\documentclass[twocolumn,preprintnumbers,amsmath,amssymb]{revtex4}

\usepackage{graphicx}
\usepackage{dcolumn}
\usepackage{bm}

\begin{document}


\title{Signatures of Quantum Transport Through Two-Dimensional Structures With Correlated and Anti-Correlated Interfaces}

\author{Tony Low$^{1}$ and Davood Ansari$^{2}$}
\affiliation{%
$^{1}$Department of Electrical and Computer Engineering. Purdue University, West Lafayette, IN 47906, USA \\
$^{2}$Department of Electrical and Computer Engineering. National University of Singapore, S119077, Singapore
}%

\date{\today}

\begin{abstract}

Electronic transport through a 2D deca-nanometer length channel with correlated and anti-correlated surfaces morphologies is studied using the Keldysh non-equilibrium Green function technique. Due to the pseudo-periodicity of these structures, the energy-resolved transmission possesses pseudo-bands and pseudo-gaps. Channels with correlated surfaces exhibit wider pseudo-bands than their anti-correlated counterparts. By surveying channels with various combinations of material parameters, we found that a smaller transport mass increases the channel transmittivity and energy bandwidth of the pseudo-bands. A larger quantization mass yields a larger transmittivity in channels with anti-correlated surfaces. For channels with correlated surfaces, the dependence of transmittivity on quantization mass is complicated by odd-to-even mode transitions. An enhanced threshold energy in the energy-resolved transmission can also be observed in the presence of surface roughness. The computed enhanced threshold energy was able to achieve agreement with the experimental data for Si$\left\langle 110\right\rangle$ and Si$\left\langle 100\right\rangle$ devices.
\end{abstract}

\maketitle

\section{\label{sec:level1}Introduction}

In the literature, theoretical studies of the physics of surface roughness on electronic transport properties mainly focus on the linear response near thermodynamic equilibrium. In this regime, transport is diffusive and the electron dynamics is well described by a Boltzmann equation \cite{fischetti92} or Kubo formula \cite{ferry99}. Once the pertubation Hamiltonian for surface roughness ($H_{SR}$) is formulated, the transition amplitude between electronic states can be computed through Fermi's golden rule. Surface roughness limited mobility can then be systematically calculated. The theory on the form of $H_{SR}$ traces back to the work by Prange and Nee \cite{prange68} on magnetic surface states in metals. More recently, a systematic derivation of $H_{SR}$ was discussed by Ando \cite{ando77} in the context of electronic transport in semiconductors. 

It is well understood that this perturbation Hamiltonian consists of two parts \cite{mou00}: (i) local energy level fluctuations and (ii) local charge density fluctuations. The first term (i.e. local energy level fluctuations) is usually introduced phenomenologically \cite{sakaki87} and explains the experimental observation that the surface roughness limited electron mobility in a quantum well scales with the film thickness ($T_{b}$) and with the material quantization mass ($m_{z}$) according to $T_{b}^{-6}$ and $m_{z}^{2}$ respectively \cite{sakaki87,uchida02,low04,tsuitsui05}. The latter term (i.e. local charge density fluctuations) is believed to be an important contribution to the degradation of electron mobility in the high inversion charge density regime \cite{jin07}. Although the treatment of the surface roughness problem is usually conducted in the framework of effective mass theory, a microscopic and self-consistent determination of $H_{SR}$ can be obtained through density functional theory \cite{evans05}. Another manifestation of surface roughness in quantum wells is the enhanced threshold energy, which has recently been observed experimentally \cite{uchida02} in quantum wells with thickness $<$$4nm$. These experiments show that the observed threshold energy does not follows the expected inverse quadratic scaling relationship with $T_{b}$. An objective of this paper is to explain why this deviation from quadratic scaling occurs.

The physical effects of surface roughness on phase coherent transport become very convoluted when the quantum well surfaces are roughened with random inhomogenuity of different scales \cite{wang05}. We limit our study to phase coherent electronic transport through quantum wells with two special kinds of surface roughness morphology: (i) perfectly correlated surface roughness and (ii) perfectly anti-correlated surface roughness morphologies. Phase coherent transport could be possible as devices are scaled into the deca-nanometer regimes \cite{schliemann04}. In practise, one would expect a quantum well grown using the atomic layer deposition technique to produce a high degree of correlation between the two surfaces. Studies of surface roughness scattering in the diffusive regime usually ignore such surface correlation effects \cite{mou00}. Herein, we show that phase coherent transport through quantum wells with perfectly correlated or perfectly anti-correlated surfaces gives rise to distinctive features in the energy resolved transmission profile. The theoretical method we had employed is the Keldysh non-equilirium Green function (NEGF) approach \cite{keldysh64,datta97,haug96,mahan90} within a finite element and boundary element discretization scheme \cite{mohan02,polizzi03,havu04}.


 This paper is organised as follows. Section II discusses the NEGF formalism and methodology in a finite element discretization scheme. Section III examines the energy-resolved transmission characteristics for quantum films with correlated and anti-correlated surfaces. We discuss these results in comparison to the Kronig Penny model \cite{buczko00}. Section IV studies the impact of quantization and transport masses on the transmission characteristics. Finally, we compare our results to experimental values of the enhanced threshold energy for Si$\left\langle 100\right\rangle$ and Si$\left\langle 110\right\rangle$ devices.

\section{\label{sec:level2}Theory and Model}


The Landauer approach \cite{landauer57,landauer70} pictures a device within which dissipative processes is absent but coupled to perfect thermodynamic systems known as `contacts'. This approach has been very successful in modeling physical effects in myriad of problems in the field of mesoscopic physics \cite{imry02}. When irreversible or energy dissipating processes are present in the device, a more sophisticated quantum transport model such as the non-equilibrium Green function (NEGF) is needed to account for the coupling and transitions between the different quantum states in the system. The NEGF method was first formulated by Kadanoff and Baym \cite{kadanoff62} and Keldysh \cite{keldysh64}. Discussions of NEGF and its applications to condensed matter phenomena can be found in textbooks by Datta \cite{datta97}, Haug \cite{haug96} and Mahan \cite{mahan90}. 

This section summarizes the NEGF formalism applied to electronic transport through a 2D channel implemented using the finite element (FEM) and boundary element (BEM) discretization scheme \cite{mohan02}. Our choice of FEM over finite difference method is mainly because it can resolve the device's roughened surface geometry more efficiently with its flexible mesh. Our methods are similar to the ones developed by Havu \itshape et. al \normalfont \cite{havu04,havu06} and Polizzi \itshape et. al \normalfont \cite{polizzi03}. 


\begin{figure}[htps]
\centering
\scalebox{0.29}[0.29]{\includegraphics*[viewport=5 115 830 550]{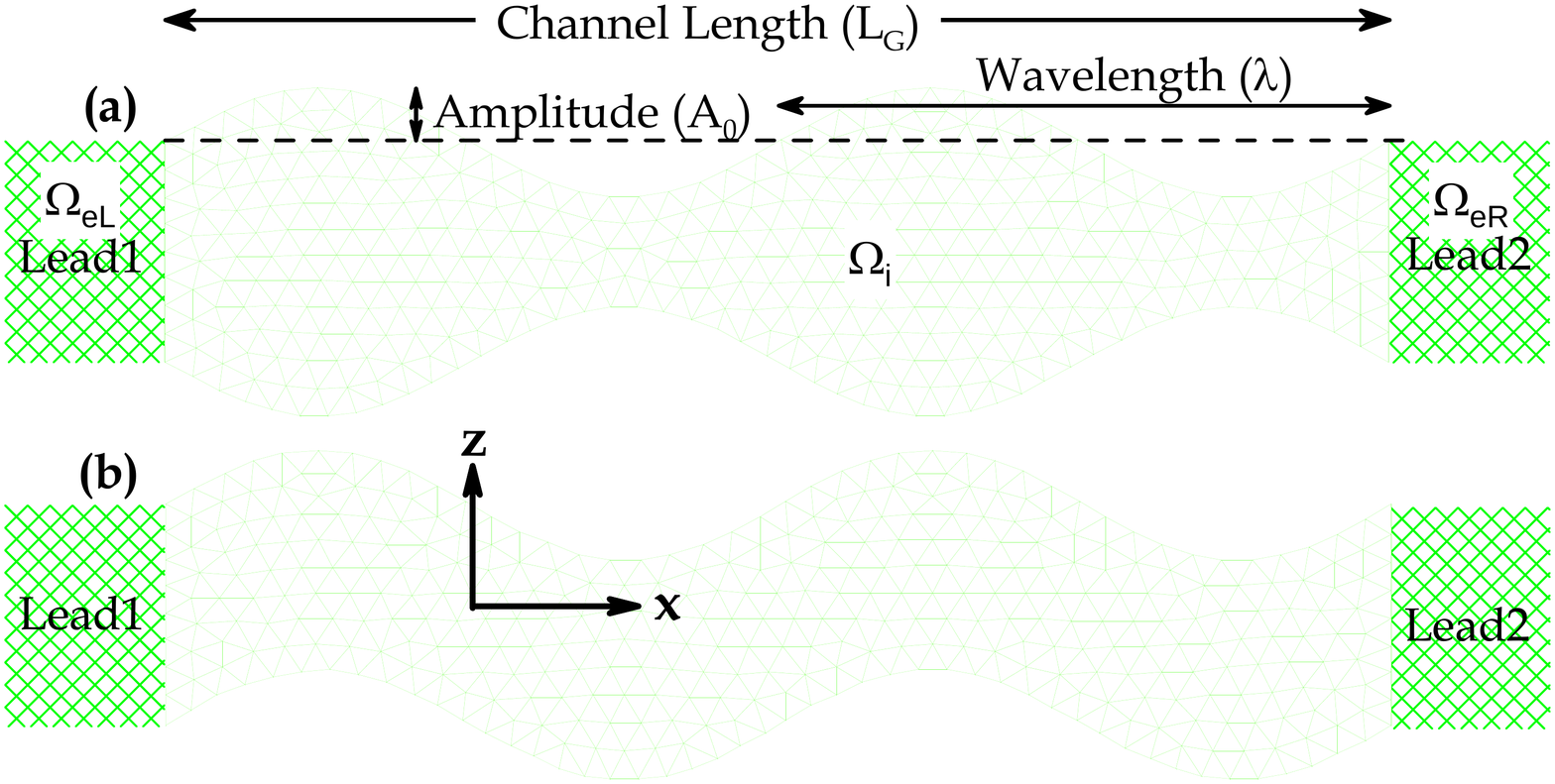}}
\caption{\footnotesize\slshape\sffamily Illustration of the mesh used for simulation of $10nm$ $2D$ channel with $\bold{(a)}$ anti-correlated and $\bold{(b)}$ perfectly correlated surfaces. In our work, the roughness is characterized by only two parameters: amplitude $(A_{0})$ and wavelength $(\lambda)$ as depicted in $\bold{(a)}$. The meshes are generated with average distance of $0.25nm$. }
\label{mesh}
\end{figure}

Fig. \ref{mesh} illustrates 2D channels with correlated and anti-correlated surface morphologies. The problem domain is denoted by $\Omega$, with points represented by a 2D spatial coordinate $r$=$(x,z)\in\Omega$. $\Omega$ is then partition into the interior domain $\Omega_i$ and exterior domains $\Omega_{ej}$, where $j$=$L,R,0$. $\Omega_{eL}$ and $\Omega_{eR}$ denotes the left and right leads respectively, while $\Omega_{e0}$ denotes the remaining space. Each exterior domain $\Omega_{ej}$ shares the boundary with $\Omega_{i}$ denoted as $\partial\Omega_{ij}$. The boundary of $\Omega_{i}$ is simply $\partial\Omega_{i}=\Sigma_{j}\partial\Omega_{ij}$. The goal is to seek the numerical solution of the Green function in $\Omega_{i}$, denoted by $G(r,r')$. In our problem, the exterior domains $\Omega_{ej}$ consists of semi-infinite leads with known Green functions \cite{datta97}. Therefore, BEM can be applied to each of these exterior domain to account for its effect on the respective boundaries $\Omega_{ij}$. The purpose of FEM is to formulate the differential equation within $\Omega_i$. \\

Within the effective mass approximation, the $2D$ Hamiltonian that we are solving can be written as,
\small
\begin{eqnarray}
\nonumber
\hat{H}\Psi(r)&\equiv&-\frac{\hbar^2}{2}\nabla_{r}\cdot\left[\bar{M}(r)\nabla_{r}\Psi(r)\right]+V(r)\Psi(r)\\
&=&\epsilon\Psi(r)
\label{s}
\end{eqnarray}
\normalsize   
where $r$=$(x,z)\in\Omega$ and $\bar{M}(r)$ is a $2$$\times$$2$ effective mass tensor. In this work, we assumed an effective mass tensor with the following matrix elements; $[\bar{M}]_{11}$$=$$m_{x}^{-1}$, $[\bar{M}]_{12}$$=$$0$, $[\bar{M}]_{21}$$=$$0$ and $[\bar{M}]_{22}$$=$$m_{z}^{-1}$. $m_{x}$ and $m_{z}$ are the transport and quantization masses respectively. The Green function of $\hat{H}$ is defined as, 
\small
\begin{eqnarray}
\left[\epsilon+i\eta-\hat{H}\right] G(r,r')=\delta(r-r') : r\in\Omega
\label{diracform}
\end{eqnarray}
\normalsize
where the boundary condition of outgoing waves is incorporated by the introduction of $\eta\rightarrow 0^{+}$, i.e. $G(r,r')$ is the retarded Green function. $\delta(r-r')$ is the Dirac delta function. In the FEM scheme, we have the node-wise shape functions $\alpha_{i}(r)$ as our basis functions \cite{mohan02} i.e. linear basis functions are employed in this work. Using the properties of a Dirac delta function of Eq. \ref{diracform}, we can write $\alpha_{h}$ as follows,
\small
\begin{eqnarray}
\nonumber
\alpha_{h}(r')
&=&\int_{r\in\Omega_{i}}\alpha_{h}(r)\left(-V(r)+\epsilon+i\eta\right)G(r,r')d\Omega\\
\nonumber
&&+\int_{r\in\Omega_{i}}\alpha_{h}(r)\frac{\hbar^2}{2}\nabla_{r}\cdot \left[\bar{M}(r)\nabla_{r} G(r,r')\right]d\Omega\\
&&: r'\in\Omega_{i}
\label{functionintegral1}
\end{eqnarray}
\normalsize   
The second integral term on the R.H.S. of Eq. \ref{functionintegral1} contains a second order differential integrand. It can be reduced to first order via the identity,
\small
\begin{eqnarray}
f_{1}\nabla \cdot \left[f_{2}\nabla f_{3}\right]=\nabla\cdot \left[f_{1}f_{2}\nabla f_{3}\right]-f_{2}\nabla f_{1}\cdot \nabla f_{3}
\end{eqnarray}
\normalsize   
Eq. \ref{functionintegral1} then becomes,
\small
\begin{eqnarray}
\nonumber
\alpha_{h}(r')&=&\int_{r\in\Omega_{i}}\alpha_{h}(r)\left(-V(r)+\epsilon+i\eta\right)G(r,r')d\Omega\\
\nonumber
&&+\int_{r\in\partial\Omega_{i}}\left(\alpha_{h}(r)\frac{\hbar^2}{2}\bar{M}(r)\nabla_{r}G(r,r')\right)\cdot \hat{n}d\partial\Omega_{i} \\
\nonumber
&&-\int_{r\in\Omega_{i}}\bar{M}(r)\nabla_{r}\alpha_{h}(r)\frac{\hbar^2}{2}\cdot\nabla_{r}G(r,r')d\Omega\\
&&: r'\in\Omega_{i}
\label{femeq1}
\end{eqnarray}
\normalsize   
where $\hat{n}$ in Eq. \ref{femeq1} is the normal vector to the boundary $\partial\Omega_{i}$. We note that in order to obtain $\nabla_{r}G(r,r')$ along the boundary $\partial\Omega_{i}$, one requires information about the Green function outside and within $\Omega_{i}$. Recall that the Green function of the exterior domain $\Omega_{ek}$, $G_{ek}$, has a simple analytical form \cite{datta97}. For our purpose, we only need to know the explicit form of $G_{ek}$ along the boundary $r_{ek}$$\in$$\partial \Omega_{ik}$ ;
\begin{eqnarray}
&&G_{ek}(x_{ek},z_{ek};x_{ek},z'_{ek})\\
\nonumber
&=&-\sum_{m=1}^{\infty}\chi_{m}(z_{ek})\chi_{m}(z'_{ek}) \frac{2sin(\omega_{m}x_{ek})exp(i\omega_{m}x_{ek})}{\hbar v_{m}}
\label{leadgf}
\end{eqnarray}
where $v_{m}$ is the carrier velocity defined as $v_{m}$=$\hbar\omega_{m}/m_{x}$ and  $\omega_{m}$=$(1/\hbar)\sqrt{2m_{x}(\epsilon-\kappa_{m})}$. $\chi_{m}$ is the eigenstates of the confined modes in the leads corresponding to the eigenstate with energy $\kappa_{m}$. In order to incorporate the information of the exterior Green function into Eq. \ref{femeq1}, we express the term $\nabla_{r}G(r,r')$ in the last integral expression as \cite{havu04},
\small
\begin{eqnarray}
\nonumber
&&\nabla_{r}G(r,r')\\
\nonumber
&=&\int_{r_{ej}\in\partial\Omega_{ij}}G(r_{ej},r')\frac{\hbar^2}{2}\bar{M}(r_{ej})\nabla_{r}\nabla_{r_{ej}} G_{ej}(r_{ej},r)d\partial\Omega_{ej}\\
&&: r\in\Omega_{ej}
\label{s}
\end{eqnarray}
\normalsize 
With this replacement to Eq. \ref{femeq1}, the calculus part of the problem is complete and we are ready to formulate the problem in matrix form.

If we assume that the Green function can be expressed in terms of the FEM basis as follows;
\small
\begin{eqnarray}
G(r,r')&\approx &\sum_{ij}\alpha_{i}(r)\alpha_{j}(r')G_{ij}
\label{s}
\end{eqnarray}
\normalsize   
Then Eq. \ref{femeq1} can be formulated into a compact matrix equation by multiply both sides by $\alpha_{g}(r')$ and integrate over $r'$. The matrix equation is,
\small
\begin{eqnarray}
S=\left[\epsilon S-H-\Sigma\right]GS
\end{eqnarray}
\normalsize   
where $[S]_{gh}$=$\int \alpha_{g}(r')\alpha_{h}(r')dr'$ is commonly known as the overlap matrix. The explicit form for computing the matrix elements of $\Sigma$ and $H$ are given in Eq. \ref{fembemfor1}.
\begin{widetext}
\small
\begin{eqnarray}
\nonumber
[H]_{hi}&\equiv&\left[\int_{r\in\Omega_{i}}\alpha_{h}(r)\left(-V(r)+i\eta\right)\alpha_{i}(r)-\frac{\hbar^2}{2}\bar{M}(r)\nabla_{r}\alpha_{h}(r)\cdot\nabla_{r}\alpha_{i}(r)d\Omega\right]\\
\label{fempart}
\nonumber
[\Sigma_{k}]_{hi}&\equiv&\left[\int_{r\in\partial\Omega_{ik}}\left(\alpha_{h}(r)\frac{\hbar^2}{2}\bar{M}(r)\int_{r_{ek}\in\partial\Omega_{ik}}\alpha_{i}(r_{ek})\frac{\hbar^2}{2}\bar{M}(r_{ek})\nabla_{r}\nabla_{r_{ek}} G_{ek}(r_{ek},r)d\partial\Omega_{ek}\right)\cdot \hat{n}d\partial\Omega_{i}\right]\\
\label{fembemfor1}
\end{eqnarray}
\normalsize
\end{widetext}

With all these quantities known, we are ready to compute the device observables through the Green function;
\small
\begin{eqnarray}
G(\epsilon)&=&\left[\epsilon S-H-\Sigma\right]^{-1}
\label{s}
\end{eqnarray}
\normalsize   
The device's local density-of-states is computed through the spectral function defined as,
\small
\begin{eqnarray}
A(\epsilon)=G(\epsilon)^{\dagger}\left[\Gamma_{L}(\epsilon)+\Gamma_{R}(\epsilon)\right]G(\epsilon)
\label{s}
\end{eqnarray}
\normalsize   
where $\Gamma$=$i[\Sigma-\Sigma^{\dagger}]$ are the broadening functions to be computed individually for each leads \cite{datta97}. The diagonal elements of $A(\epsilon)$ yields the local density-of-states. Finally, the transmission is computed by taking the trace of the transmission function $\Phi$ defined to be \cite{datta97},
\small
\begin{eqnarray}
\Phi(\epsilon)=\Gamma_{L}(\epsilon)G(\epsilon)\Gamma_{R}(\epsilon)G(\epsilon)^{\dagger}
\label{s}
\end{eqnarray}
\normalsize   
To facilitate our subsequent analysis, we unclustered the total transmission $\Phi$ and examine only the transmission characteristics between mode $m=1,2$ of left lead to mode $n=1,2$ of right lead. This mode-to-mode transmission ($\Phi_{mn}$) is easily accomplished in our numerical scheme by noting that the lead self-energy can be unfolded into their respective modes:
\begin{eqnarray}
G_{ek}=G^{1}_{ek}+G^{2}_{ek}+G^{3}_{ek}\ldots
\label{s}
\end{eqnarray}
where $G^{m}_{ek}$ is the self energy for mode $m$. This allows one to define a broadening function associated with each mode $\Gamma_{L/R}^{m/n}$, from which the respective mode-to-mode transmission $\Phi_{mn}$ can be computed. With this understanding, we shall begin our numerical analysis.

\begin{figure}[htps]
\centering
\scalebox{0.34}[0.34]{\includegraphics*[viewport=40 15 780 560]{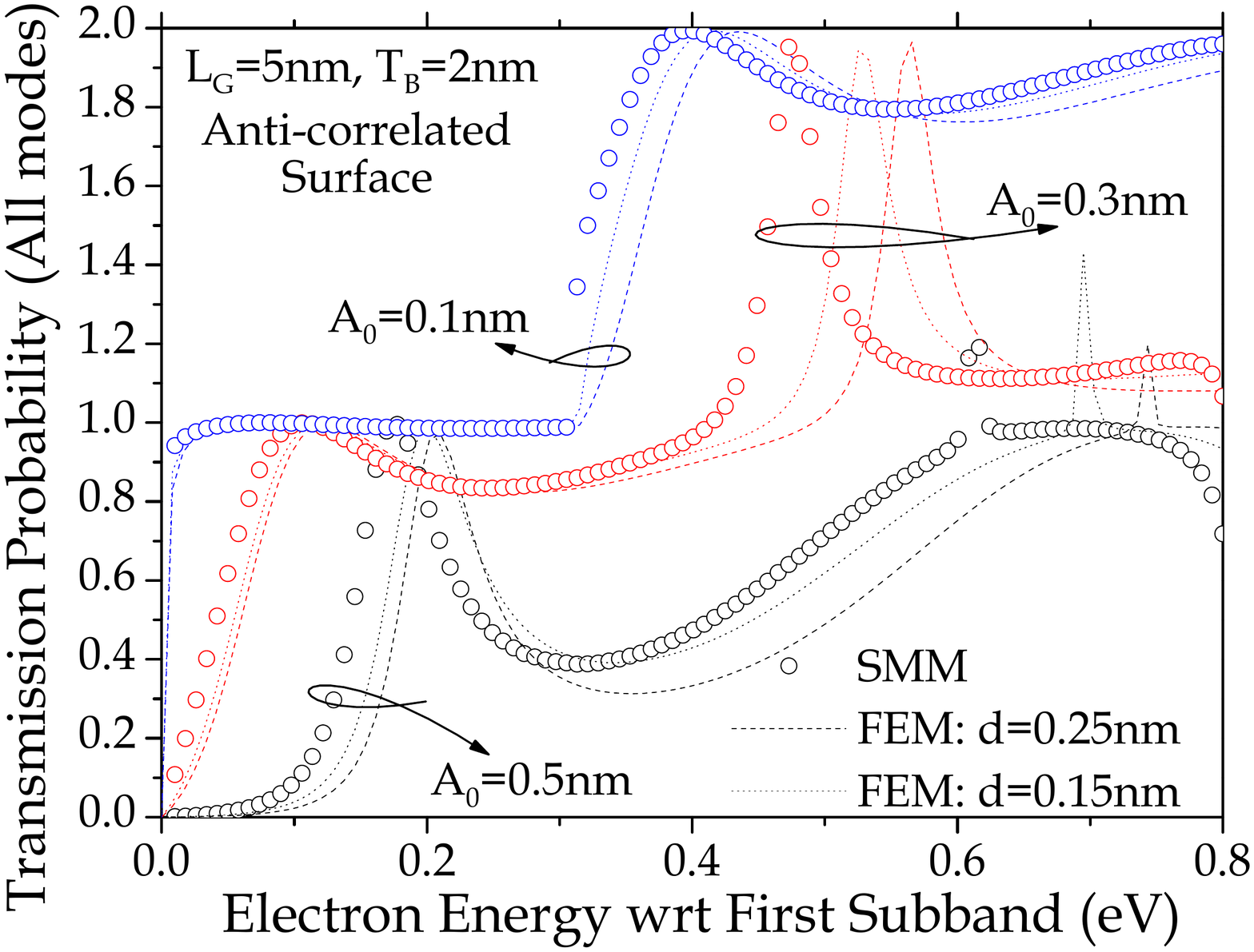}}
\caption{\footnotesize\slshape\sffamily FEM and Scattering matrix method (SMM) simulation of a $L_{G}=5nm$ and $T_{B}=2nm$ channel with anti-correlated surfaces having roughness amplitudes $A_{0}=0.1,0.3,0.5nm$ and wavelength $\lambda=2.5nm$. We assumed a transport mass $m_{x}=0.2m_{0}$ and quantization mass $m_{z}=0.9m_{0}$ i.e. based off Si material. SMM is a 'mode space' approach which the quantized energy is resolved analytically.}
\label{compareFEMSMM}
\end{figure}

Fig.\ref{mesh} illustrates a typical FEM mesh with $\bold{(a)}$ anti-correlated and $\bold{(b)}$ perfectly correlated surfaces used in our calculations. The FEM mesh is generated using the algorithm developed by Persson and Strang \cite{persson04} based on the well-known Delaunay triangulation routine. In our work, the roughness is characterized by sinusoidal profiles with only two parameters: amplitude $(A_{0})$ and wavelength $(\lambda)$. These parameters are analogous to the root-mean-square roughness and the roughness auto-correlation length commonly employed in the literatures to describe the surface morphology \cite{goodnick85}. In Fig. \ref{compareFEMSMM}, we compare the energy-resolved transmission calculated with our FEM-BEM method to the scattering matrix method (SMM) \cite{sheng97} using various mesh sizes $d$. In the SMM approach, the quantized energies are resolved analytically unlike the FEM-BEM or finite difference approaches.  The simulations used a channel length $L_{G}$=$5nm$, an average film thickness $T_{B}$=$2nm$, a transport mass $m_{x}$=$0.2m_{0}$, and quantization mass $m_{z}$=$0.9m_{0}$. Anti-correlated surfaces described by roughness amplitudes $A_{0}$=$0.1, 0.3, 0.5nm$ and wavelength $\lambda$=$2.5nm$ were considered. As shown in Fig. \ref{compareFEMSMM}, the FEM-BEM results are in satisfactory agreement with SMM. A very fine spatial discretization in the transport direction was employed for the SMM calculations.

\begin{figure}[htps]
\centering
\scalebox{0.35}[0.35]{\includegraphics*[viewport=50 27 780 585]{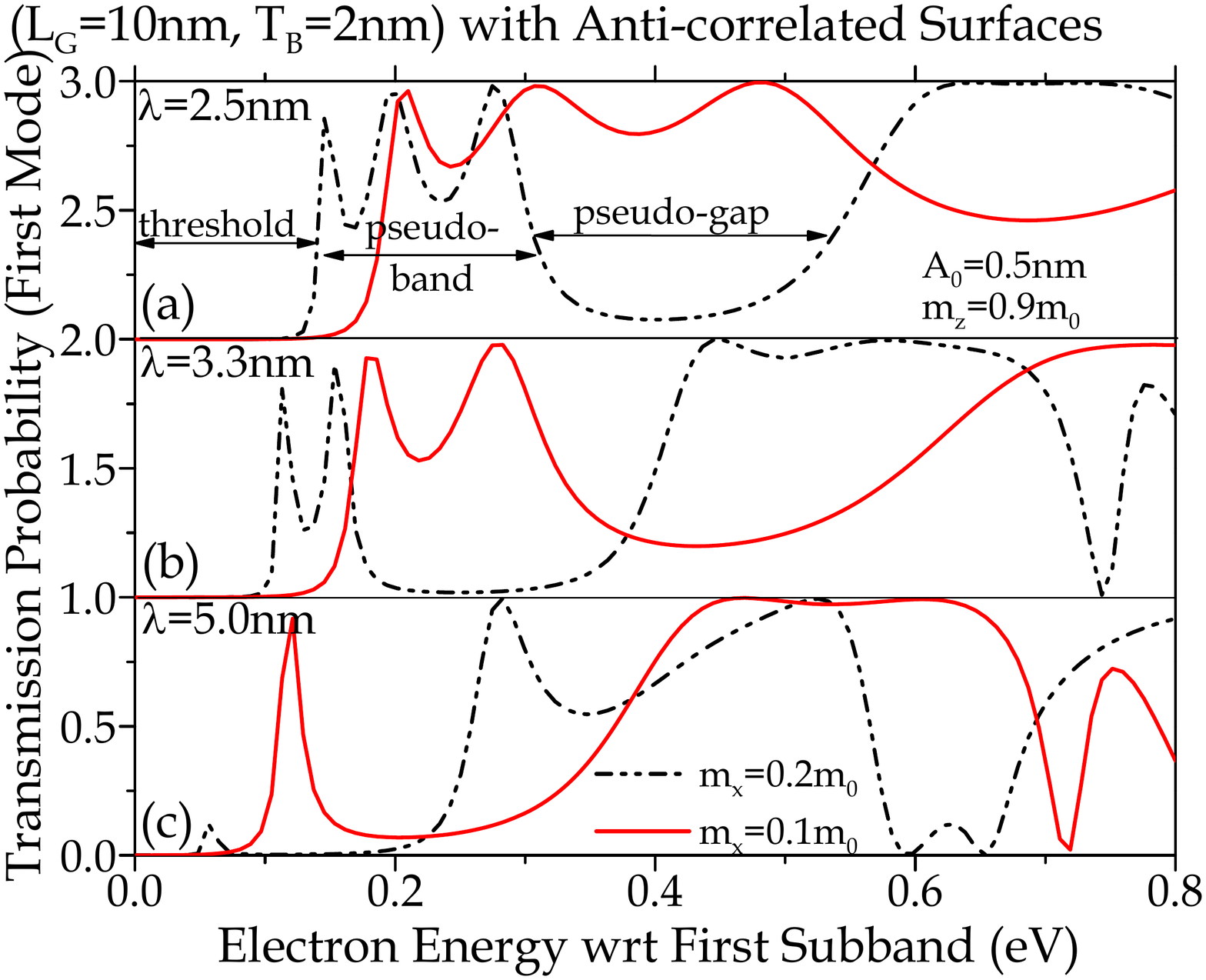}}
\caption{\footnotesize\slshape\sffamily Energy-resolved transmission (for $\Phi_{11}$) from FEM simulation of a $L_{G}=10nm$ and $T_{B}=2nm$ channel with various anti-correlated surfaces with roughness amplitudes $A_{0}=0.5nm$ and wavelength $\bold{(a)}$ $\lambda=2.5nm$, $\bold{(b)}$ $\lambda=3.3nm$ and $\bold{(c)}$ $\lambda=5.0nm$. For each structure, we simulated a transport mass of $m_{x}=0.2m_{0}$ (dash-dotted lines) and $m_{x}=0.1m_{0}$ (solid lines) each with quantization mass $m_{z}=0.9m_{0}$.}
\label{antiSRmode1}
\end{figure}
\section{\label{sec:level3}General Features in Energy Resolved Transmission}
The objective of this section is to perform a systematic analysis of the transmission characteristics of electronic transport through a $2D$ quantum channel with correlated and anti-correlated surfaces. As a model system, we adopted a set of parameter values typical of `end-of -road-map' devices \cite{itrs07} i.e. a $L_{G}$=$10nm$ $2D$ channel with an average quantum film thickness of $T_{B}$=$2nm$. For each class of devices, the effects of material properties and the degree of roughness on the energy resolved transmission characteristics were examined. The material bandstructure was parameterized by a set of effective masses i.e. transport mass $m_{x}$ and quantization mass $m_{z}$. Although microscopic details of the atomic structure of interfaces \cite{evans05} and related elemental defects (i.e. Si-Si and Si-O-Si bonds) \cite{buczko00} could be a source of interface scattering, it is not the main focus of this work. Ignoring these interfacial elemental defects, we study the geometrical effect of surface roughness on the transport characteristics. 


\subsection{\label{sec:level3}Transmission Through Anti-Correlated Surfaces}
Fig.\ref{antiSRmode1} shows the energy-resolved transmission characteristics $\Phi_{mn=11}$ (first mode to first mode transmission) for devices with anti-correlated surfaces. Three different sets of roughness parameters were simulated: (a) $A_{0}$=$0.5nm$ and $\lambda$=$2.5nm$, (b) $A_{0}$=$0.5nm$ and $\lambda$=$3.3nm$, (c) $A_{0}$=$0.5nm$ and $\lambda$=$5.0nm$. For each set of roughness parameters, the following sets of material parameters were simulated: (dash-dotted lines) $m_{z}$=$0.9m_{0}$ with $m_{x}$=$0.2m_{0}$ and (solid lines) $m_{z}$=$0.9m_{0}$ with $m_{x}$=$0.1m_{0}$. 

Several general observations can be made about the energy resolved transmission spectra: (i) there are regions of pseudo-gaps and pseudo-bands where transmissivity is relatively opaque and transparent respectively, (ii) within the pseudo-bands, there are camel-back structures which increases in number with increasing roughness frequency (i.e. decreasing $\lambda$) and (iii) a delayed `turn-on' of transmission called `enhanced threshold', which increases with increasing roughness frequency. The enhanced threshold energy is a geometrically derived property due to $2D$ quantization effects of the roughened morphology. The enhanced threshold is zero for an unroughened quantum well channel. Note that the energy scale are referenced from the subband energy of the first mode in the lead, previously defined as $\kappa_{1}$. In addition, due to the symmetry of the problem, $\Phi_{mn}$$\neq$$ 0$ if and only if both $m$ and $n$ are odd/even numbers. 


Due to the anti-correlated surface morphology, the thickness of the $2D$ quantum film fluctuates from the source to drain contacts. Henceforth, we can visualise the electron as moving across the channel through an undulating energy landscape caused by the variable film thickness. This accounts for the appearance of camel-back structures in the pseudo-band region of the energy-resolved transmission as depicted in Fig.\ref{antiSRmode1} i.e. a signature of resonant tunneling. The undulating energy landscape ($\epsilon_{QW}(x)$) can be modeled by a series of quantum wells ($E_{QW}$) as follows;
\small
\begin{eqnarray}
\nonumber
\epsilon_{QW}(x)&=&\frac{\hbar^{2}\pi^{2}}{2m_{z}\left[T_{B}+2A_{0}cos(2\pi x/\lambda)\right]^{2}}=\sum_{j}E_{QW}(x+j\lambda)\\
\label{s}
\end{eqnarray}
\normalsize 
$E_{QW}$ can be expanded as a Taylor series to give the following leading order terms:
\small
\begin{eqnarray}
\nonumber
E_{QW}(x)\approx\left[\frac{\hbar^{2}\pi^{2}}{2m_{z}(T_{B}+2A_{0})^{2}}+\frac{4\hbar^{2}\pi^{4}A_{0}x^{2}}{m_{z}(T_{B}+2A_{0})^{3}\lambda^{2}}\right]U(x)\\
\label{qws}
\end{eqnarray}
\normalsize 
where $U(x)$ is a unit pulse at $-\lambda/2$$<$$x$$<$$\lambda/2$. The second term in Eq. \ref{qws} with the kinetic operator will yield the quantized energy levels of a harmonic oscillator with energies $\xi_{n}$=$\hbar \omega(n+0.5)$ where,
\small
\begin{eqnarray}
\omega=\sqrt{\frac{8\hbar^{2}\pi^{4}A_{0}}{\lambda^{2}m_{x}m_{z}(T_{B}+2A_{0})^{3}}}
\label{omegacall}
\end{eqnarray}
\normalsize 
From the above expression, we have $\omega$$\propto$$m_{x}^{-0.5}$. Indeed, Fig.\ref{antiSRmode1} reveals that a smaller $m_{x}$ will yield a wider energy seperation between the peaks of the camel-back structure. This also translates to a larger pseudo-band bandwidth. For the structure illustrated in Fig.\ref{antiSRmode1}, the estimated first quantized energy using Eq.\ref{omegacall} is $\xi_{1}$$\approx$ $0.1357eV$, $0.1019eV$ and $0.0679eV$ for $\lambda$=$2.5nm$, $3.33nm$ and $5nm$,  respectively. These estimates are in good agreement with threshold energies computed in Fig.\ref{antiSRmode1}, which is the energy needed for the first appearance of a transmission resonance. 
\begin{figure}[t]
\centering
\scalebox{0.35}[0.35]{\includegraphics*[viewport=40 20 710 550]{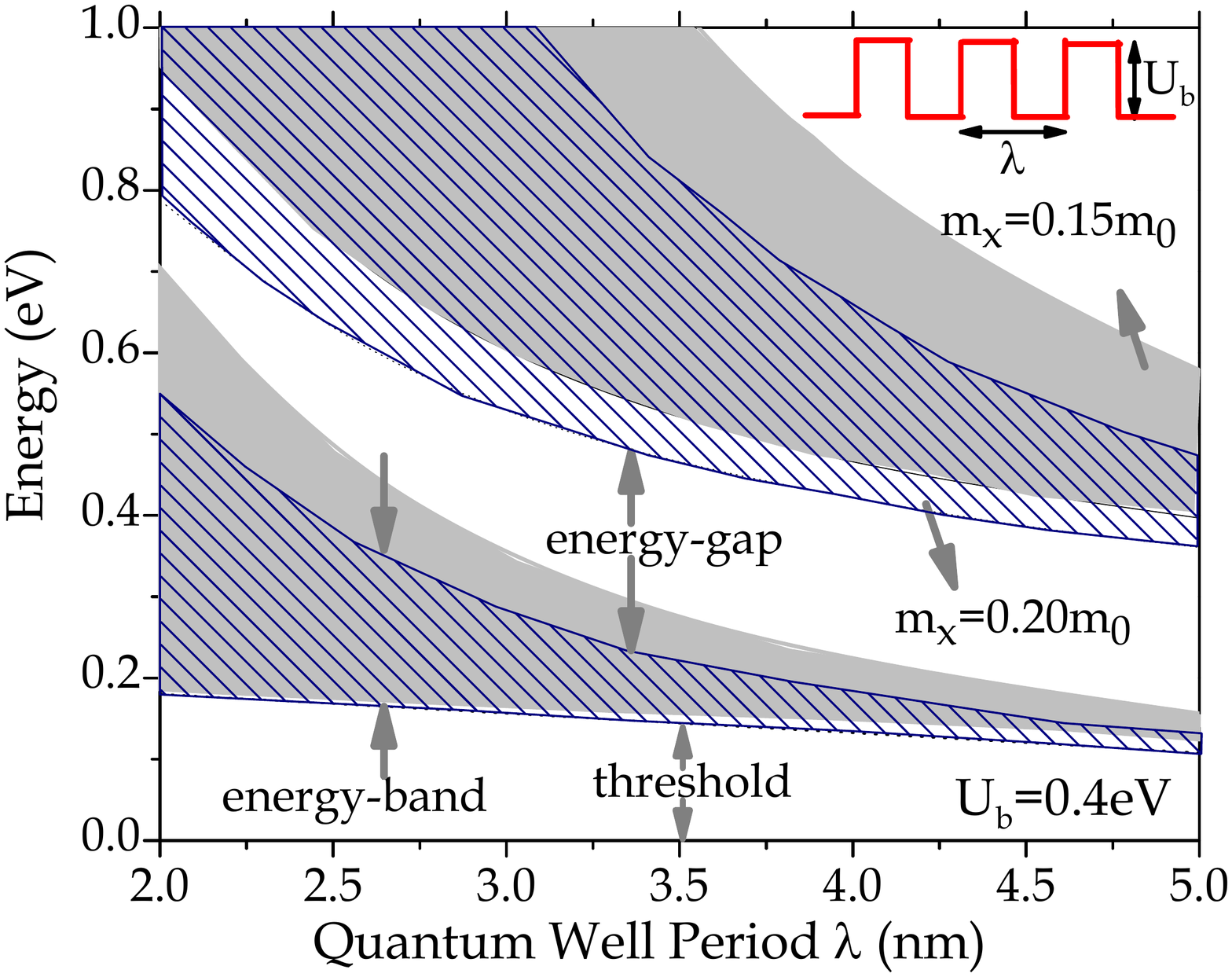}}
\caption{\footnotesize\slshape\sffamily Available energy band states (indicated by shaded regions) for a Kronig Penny 1D crystal with periodically varying square potential wells plotted with respect to the quantum well width $\lambda$. The quantum well energy barrier $U_{b}$ is set to be $0.4eV$. Effective mass of $m_{x}=0.15m_{0}$ (solid) and $m_{x}=0.2m_{0}$ (patterned) are considered.}
\label{kronig}
\end{figure}
\begin{figure}[t]
\centering
\scalebox{0.35}[0.35]{\includegraphics*[viewport=50 20 780 575]{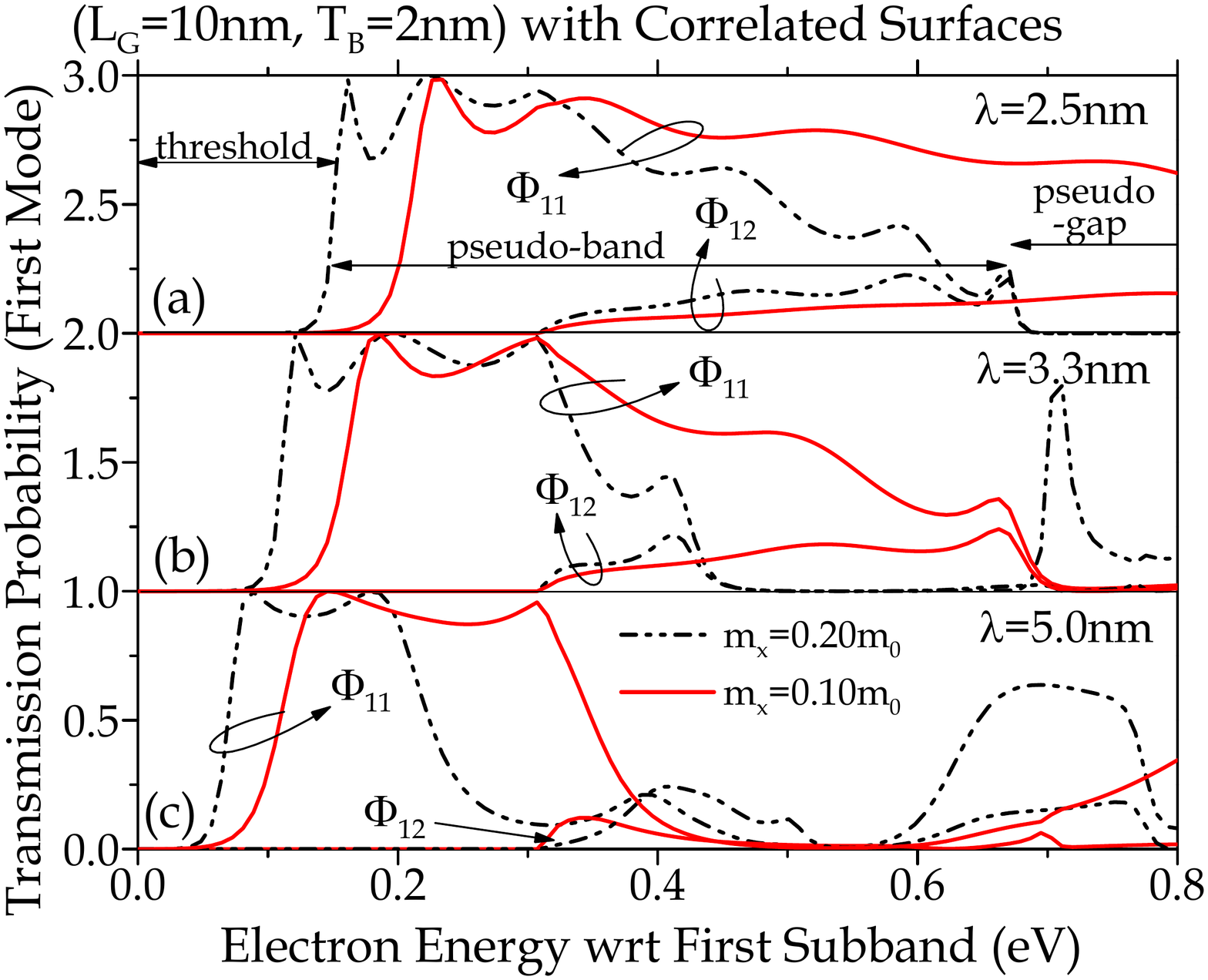}}
\caption{\footnotesize\slshape\sffamily Energy-resolved transmission (for $\Phi_{11}$ and $\Phi_{12}$) from FEM simulation of a $L_{G}=10nm$ and $T_{B}=2nm$ channel with various perfectly correlated surfaces with roughness amplitudes $A_{0}=0.5nm$ and wavelength $\bold{(a)}$ $\lambda=2.5nm$, $\bold{(b)}$ $\lambda=3.3nm$ and $\bold{(c)}$ $\lambda=5.0nm$. For each structure, we simulated a transport mass of $m_{x}=0.2m_{0}$ (dash-dotted lines) and $m_{x}=0.1m_{0}$ (solid lines) each with quantization mass $m_{z}=0.9m_{0}$. }
\label{perfectSRmode1}
\end{figure}

To explain the global features of pseudo-bands and pseudo-gaps in the energy-resolved transmission results, we shall first review the pertinent results from the Kronig-Penny model \cite{kronig31}. Consider a periodically varying rectangular energy barrier as depicted in the inset of Fig. \ref{kronig} with barrier energy of $U_{b}$ and a spatial period of $\lambda$. The bandstructure (energy $\xi$ versus momentum $\kappa$ dispersion relation) of this $1D$ periodic potential can be described by the following transcendental equation;
\small
\begin{eqnarray}
\kappa &=&\pm\frac{1}{\lambda}tan^{-1}\left(\frac{\sqrt{4-\alpha_{i}^{2}}}{\alpha}\right)
\label{s}
\end{eqnarray}
\normalsize 
where $\alpha$ is given by;
\small
\begin{eqnarray}
\nonumber
\alpha &=& 2cosh(\frac{k_{2}\lambda}{2})cos(\frac{k_{1}\lambda}{2})\\
&& +\frac{k_{2}^{2}-k_{1}^{2}}{k_{1}k_{2}}sinh(\frac{k_{2}\lambda}{2})sin(\frac{k_{1}\lambda}{2})
\label{s}
\end{eqnarray}
\normalsize 
when $0$$<$$\xi$$<$$U_{b}$ and 
\small
\begin{eqnarray}
\nonumber
\alpha &=&  2cos(\frac{k_{2}\lambda}{2})cos(\frac{k_{1}\lambda}{2})\\
&&-\frac{k_{2}^{2}+k_{1}^{2}}{k_{1}k_{2}}sin(\frac{k_{2}\lambda}{2})sin(\frac{k_{1}\lambda}{2})
\label{s}
\end{eqnarray}
\normalsize 
when $\xi$$>$$U_{b}$. In addition, we have $\hbar k_{1}$=$\sqrt{2m_{x}|\xi|}$ and $\hbar k_{2}$=$\sqrt{2m_{x}|\xi-U_{b}|}$.

Fig. \ref{kronig} surveys the bandstructure of a Kronig Penny 1D crystal under different quantum well period $\lambda$ where regions with propagating states are shaded. The analysis set is conducted for transport masses of $m_{x}$=$0.15m_{0}$ (colored region) and $m_{x}$=$0.2m_{0}$ (shaded region). The following observations can be made: (i) the energy bandwidth and threshold energy increase monotically with decreasing quantum well period $\lambda$, (ii) energy bandwidth decreases monotonically with decreasing quantum well period $\lambda$ and (iii) a smaller transport mass $m_{x}$ yields a larger energy bandwidth. Examination of the energy-resolved transmission in Fig.\ref{antiSRmode1} shows that the pseudo-band and enhanced threshold in the quasi-periodic $2D$ structure with anti-correlated surface roughness exhibits a strikingly similar trend with the Kronig Penny model analysis. Based on these arguments, we conclude that the generic features of pseudo-band and pseudo-gap observed in the energy-resolved transmission of the $2D$ film with anti-correlated surfaces is a result of the film's quasi-periodicity. 
 
\begin{figure}[t]
\centering
\scalebox{0.5}[0.5]{\includegraphics*[viewport=10 140 650 540]{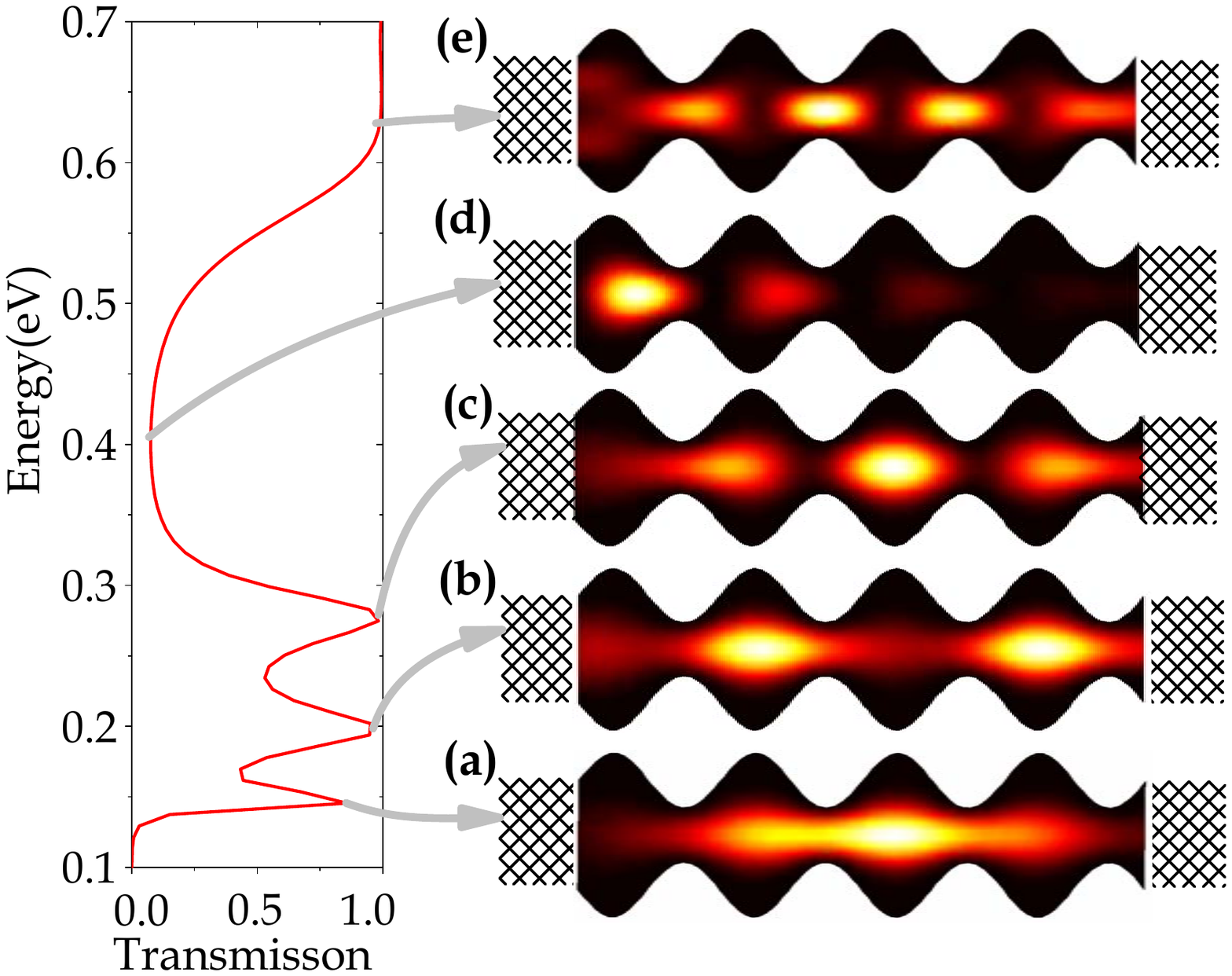}}
\caption{\footnotesize\slshape\sffamily Intensity plot of local density of states due to the first mode of left contact i.e. $G^{\dagger}\Gamma_{L}^{1}G$, each at different injection energies (measured with respect to the lead's first quantized mode) as indicated. Device structure used is a $L_{G}=10nm$ and $T_{B}=2nm$ channel with anti-correlated surfaces with roughness amplitudes $A_{0}=0.5nm$ and wavelength $\lambda=2.5nm$. A transport mass of $m_{x}=0.2m_{0}$ and quantization mass $m_{z}=0.9m_{0}$ are used. The corresponding energy-resolved transmission characteristics is plotted on the left.}
\label{resonanceLDOSa}
\end{figure}
\begin{figure}[t]
\centering
\scalebox{0.5}[0.5]{\includegraphics*[viewport=10 140 650 580]{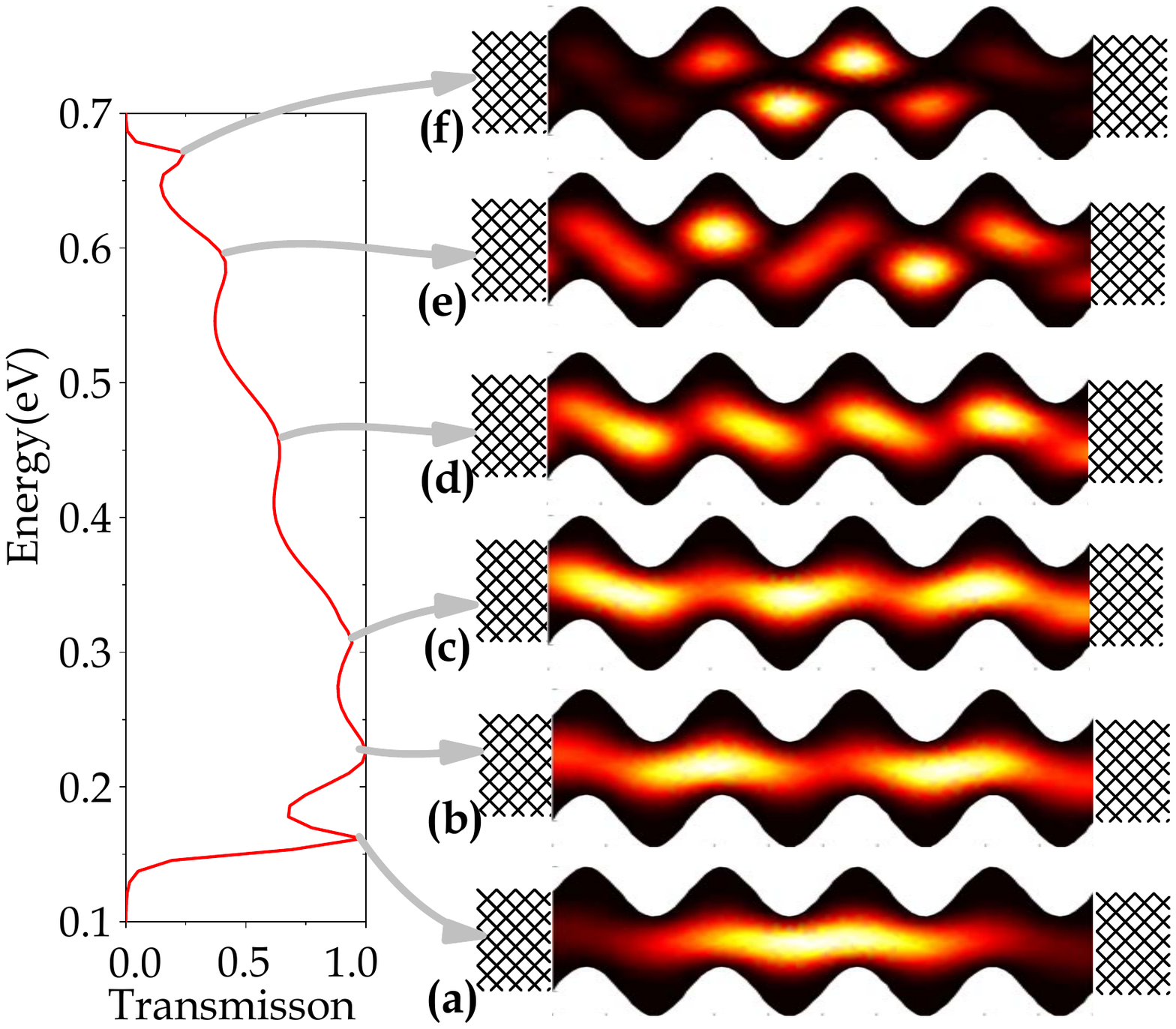}}
\caption{\footnotesize\slshape\sffamily Similar to Fig.\ref{resonanceLDOSa}, except for a channel with perfectly-correlated surfaces.  }
\label{resonanceLDOSp}
\end{figure}

\subsection{\label{sec:level3}Transmission Through Perfectly-Correlated Surfaces}

Fig. \ref{perfectSRmode1} shows the energy-resolved transmission characteristic of a $2D$ channel with perfectly correlated surface roughness. The same sets of devices as performed in the anti-correlated case, with various surface roughness parameters and transport mass $m_{x}$, are simulated. Due to the different symmetry in this case, $\Phi_{mn}\neq 0$ for odd to even mode transition, unlike the situation for anti-correlated surfaces. Therefore, we plotted the transmission characteristics $\Phi_{11}$ and $\Phi_{12}$ in Fig. \ref{perfectSRmode1}, although we will mainly focus on $\Phi_{11}$ in the subsequent disussion. 

As shown in Fig. \ref{perfectSRmode1}, the generic features of pseudo-band, pseudo-gap and enhanced threshold, induced by the quasi-periodicity are also present in structures with perfectly correlated surfaces. For each particular set of surface roughness parameters, the channel with perfectly correlated surfaces exhibits a distinctively larger pseudo-band than its anti-correlated counterpart. Since there is no variations of quantum well thickness along the channel, the scattering in this case is purely a result of $2D$ geometrical and side-wall boundary effects. As a result, one would expect the carrier to feel a less undulating energy landscape in the perfectly correlated case. Effectively, this translates to a smaller $U_{b}$ in the Kronig-Penny picture, which will then correpondingly yield a larger pseudo-band bandwidth.

Fig. \ref{resonanceLDOSa} and \ref{resonanceLDOSp} are intensity plots for the local density-of-states $G^{\dagger}(\epsilon)\Gamma_{L}^{1}(\epsilon) G(\epsilon)$ due to injection of carriers from the source contact for the anti-correlated and correlated cases respectively. Brighter regions indicate higher density-of-states. The carriers are injected from the first mode eigenstates of the source lead. The surface roughness morphologies are both characterized by $A_{0}$=$0.5nm$ and $\lambda$=$2.5nm$. The energy resolved transmission spectra are plotted on their left. The local density-of-states at each of the resonance energies reveals localised high intensity patterns, which has its origin in interference effects due to multiple scattering of waves. One observes that the number of localised spots increases with the index of the resonance level. In the energy-resolved transmission, twice the number of resonance energy levels in the pseudo-band is observed in the perfectly correlated case as compared to anti-correlated case. This is attributed to the reduced symmetry of the channel with a perfectly-correlated surface morphology, which lacks the x-axis mirror symmetry of the anti-correlated surface. The degradation of the energy-resolved transmission at higher energy is due to the appearance of the second mode at $~0.3eV$, which opens up a transmission through $\Phi_{12}$. Therefore the resonance peak has maximum transmission less than $1$. 

\begin{figure*}[t]
\centering
\scalebox{1.3}[1.3]{\includegraphics*[viewport=20 21 450 140]{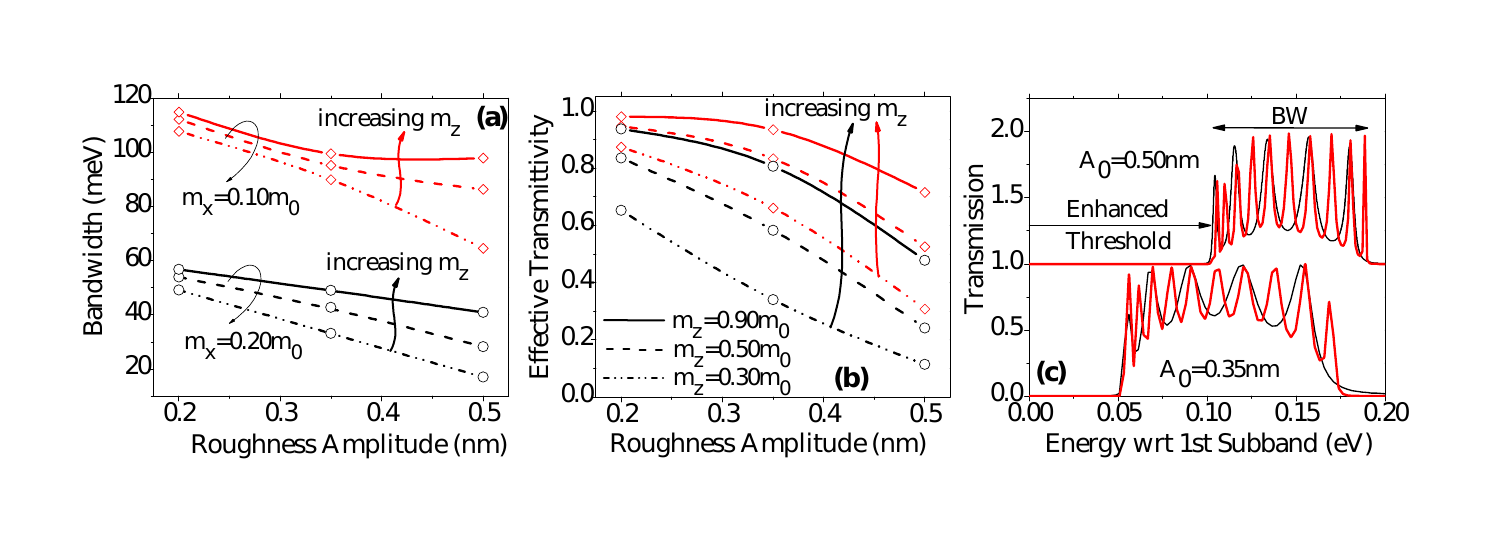}}
\caption{\footnotesize\slshape\sffamily Surverying of pseudo-band's bandwidth of the energy-resolved transmission for $\Phi_{11}$ (see $\bold{(a)}$) and the transmittivity (see $\bold{(b)}$) for various transport masses ($m_{x}=0.1m_{0}$ and $0.2m_{0}$, denoted by red and black lines respectively) and quantization masses ($m_{z}=0.3m_{0},0.5m_{0}$ and $0.9m_{0}$, denoted by dash-dotted, dashed and solid lines respectively), plotted as a function of roughness amplitudes $A_{0}$. Simulated for a device with $L_{G}=10nm$ and an average $T_{B}=2nm$ channel with anti-correlated surface roughness morphology ($\lambda=3.3nm$). $\bold{(c)}$ depicts the energy-resolved transmission for similar devices but with $L_{G}=20nm$ (black) and $40mm$ (red), with various $A_{0}$=$0.35nm$ and $0.50nm$ using effective masses of $m_{x}=0.2m_{0}$ and $m_{z}=0.9m_{0}$.}
\label{AnalysisAnti2}
\end{figure*}
\begin{figure}[t]
\centering
\scalebox{0.52}[0.52]{\includegraphics*[viewport=10 41 400 540]{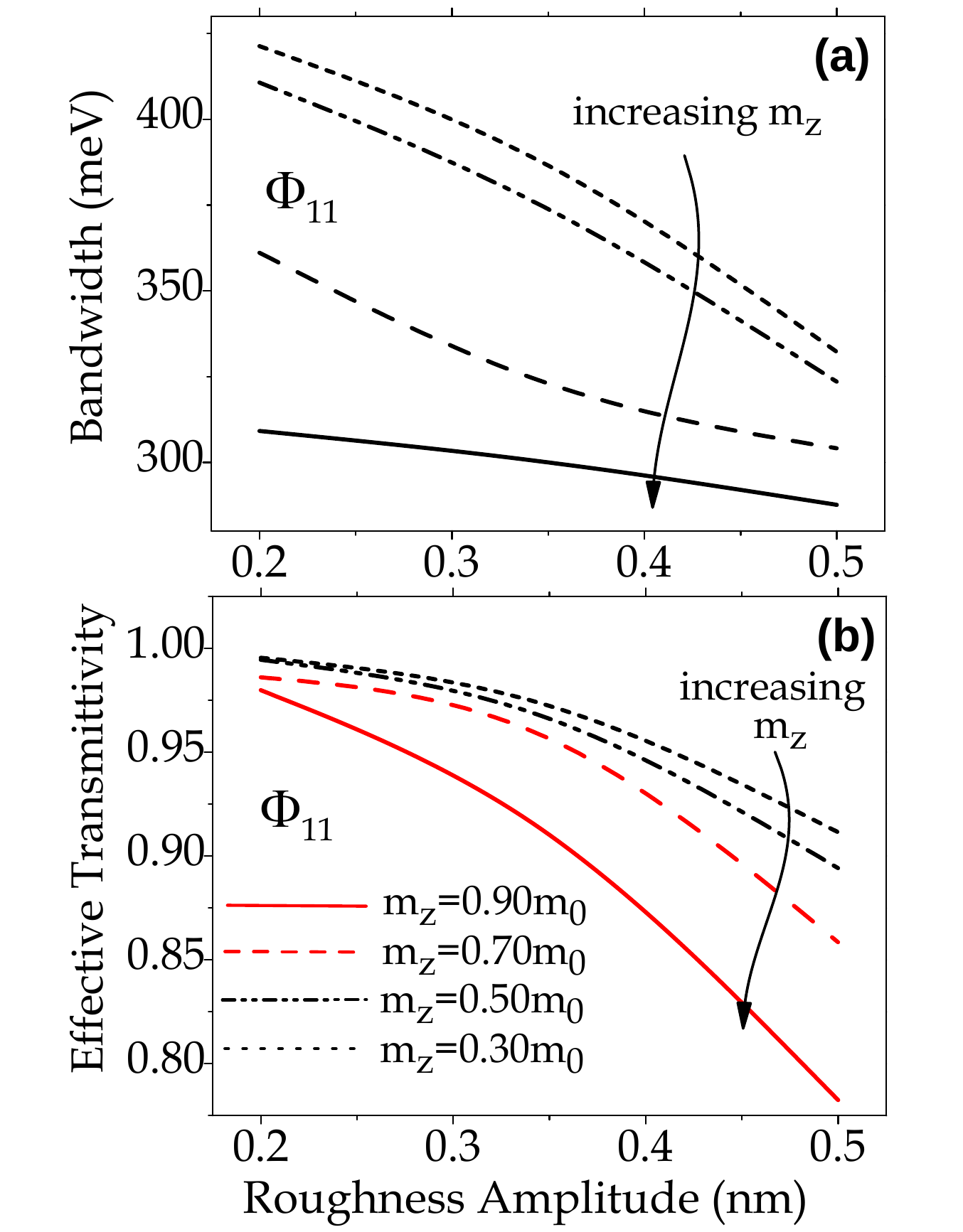}}
\caption{\footnotesize\slshape\sffamily Effects of quantization mass ($m_{z}$) on the pseudo-band's bandwidth of the energy-resolved transmission for $\Phi_{11}$ (see $\bold{(a)}$) and the transmittivity (see $\bold{(b)}$) through a channel with perfectly-correlated surface roughness morphology plotted as a function of $A_{0}$. The device has $L_{G}=10nm$ and an average $T_{B}=2nm$ channel. Transport mass of $m_{x}=0.2m_{0}$ is used for all curves. The red curves in $\bold{(b)}$ denotes that higher modes transitions, i.e. $\Phi_{12}$ and $\Phi_{22}$, are present within the pseudo-bands of these structures. }
\label{AnalysisPerfect}
\end{figure}
\section{\label{sec:level2}Effective Masses on Transmittivity and Threshold Energy}
In this section, we study the impact of transport and quantization masses and surface roughness morphologies on the general characteristics of the energy-resolved transmission through a deca-nanometer channel. Our analysis will be confined to the study of the first mode to first mode transmission, $\Phi_{11}$. We begin by proposing a reasonable metric for the measure of the transmittivity $K$ of a channel:
\small
\begin{eqnarray}
K=\frac{1}{BW}\int_{BW}\Phi(\epsilon_{xz})d\epsilon_{xz}
\label{omegacal}
\end{eqnarray}
\normalsize 
where the pseudo-band bandwidth ($BW$) is defined as the energy difference between the last and first resonance peaks in the pseudo-band. Enhanced threshold energy is defined as the energy for the appearance of the first resonance peak with respect to the lead's first subband energy, i.e. $\hbar^{2}\pi^{2}/(2m_{z}T_{B}^{2})$. This phenomenon has been observed experimentally \cite{tsuitsui05,uchida02} and the suppression of its effect is pertinent to electronic device applications. Fig. \ref{AnalysisAnti2}(c) provides an illustration of the concept of BW and enhanced threshold energy. 

Fig. \ref{AnalysisAnti2}(a)(b) and \ref{AnalysisPerfect}(a)(b) surveys the $BW$ and $K$ for a deca-nanometer channel of $\left\langle T_{B}\right\rangle$=$2nm$ with anti-correlated and perfectly correlated surface roughness morphologies respectively. Different sets of material parameters, i.e. $m_{x,z}$, are employed in the study. The key results can be summarized as follows: (i) a smaller $m_{x}$ improves the pseudo-band's bandwidth ($BW$) and effective transmittivity ($K$) for channels with anti-correlated and perfectly correlated surface roughness morphology, (ii) a larger $m_{z}$ improves the transmittivity for channels with anti-correlated surface roughness but slightly degrades the transmittivity for channels with perfectly correlated surface roughness. The observation (i) is due to the increase of resonance linewidth as derived from the higher tunneling probability due to a smaller $m_{x}$. 

In the diffusive regime, the surface roughness limited mobility for a quantum well scales proportionally with $\approx m_{z}^{2}\Delta^{-2}$, where $\Delta$ is the root-mean-square averaged fluctuation of the quantum well thickness \cite{low04}. In the channel with anti-correlated surfaces, $\Delta$=$\sqrt{2}A_{0}$. As depicted in Fig. \ref{AnalysisAnti2}(b), $K$ increases with increasing $m_{z}$. For the channel with perfectly correlated surface morphology, $\Delta$ is zero. Thus, the main source of scattering mechanism in the `classical' sense is atrributed to a local fluctuation of wavefunction \cite{mou00}. In general, a larger $m_{z}$ `propagates' the electron closer to the surfaces and renders it more sensitive to the surface roughness morphology. This explains the small degradation of $K$ with increasing $m_{z}$. However, we must emphasize that the larger degradation of $K$ (red curves in \ref{AnalysisPerfect}(b)) for $m_{z}=0.7m_{0},0.9m_{0}$ is due to the appearance of the second mode, leading to a degradation of $\Phi_{11}$ while $\Phi_{12}$ begins to increase. 

With a larger transmission bandwidth and relatively weak dependence of transmittivity on $m_{z}$, channel with highly correlated surfaces is more optimal for electronic transport in the phase coherent regime. Especially for a channel with small quantization mass, i.e. III-V semiconductor alloys, the latter property is highly desireable for ballistic transport. Although these studies are conducted for a deca-nanometer channel, we expect the results to be consistent for longer channel. To confirm this proposition, we considered the energy-resolved transmission for devices with $L_{G}$=$20nm$ and $40nm$ in Fig. \ref{AnalysisAnti2}(c). The increase in channel length results in more resonance peaks (i.e. the number of peaks within the pseudo-band is proportional to the number of sinusoidal cycles in the roughness morphology of the channel) but with the global features of enhanced threshold and the pseudo-band's bandwidth intact. In particular, its energy-resolved transmission exhibits similar enveloping characteristics for different $L_{G}$, where the envelope is characteristics of a given $A_{0}$. 

\begin{figure}[t]
\centering
\scalebox{0.43}[0.43]{\includegraphics*[viewport=150 58 750 542]{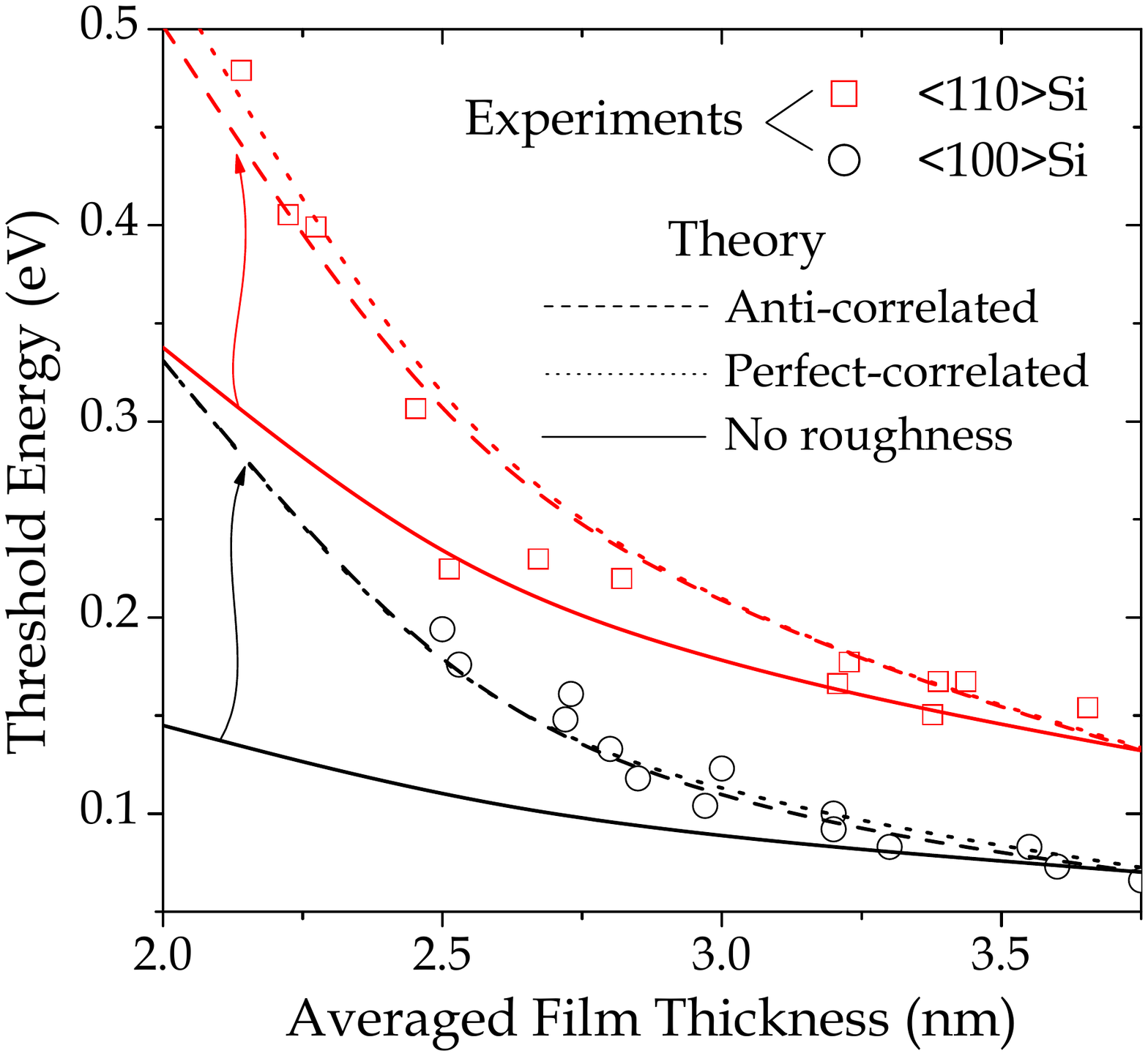}}
\caption{\footnotesize\slshape\sffamily Theoretically calculated threshold energy as a function of the averaged 2D quantum film's thickness compared with the experimental data for a Si$\left\langle 110\right\rangle$ \cite{tsuitsui05} and Si$\left\langle 100\right\rangle$ \cite{uchida02} devices. For our calculations, a roughness amplitude of $A_{0}=0.5nm$ and $A_{0}=0.6nm$ are able to describe the experimental data for Si$\left\langle 110\right\rangle$ and Si$\left\langle 100\right\rangle$ devices respectively, where a roughness wavelength of $\lambda=2.5nm$ was assumed for both cases. The threshold energy for a channel with unroughened (solid lines), anti-correlated (dashed lines) and perfectly correlated (dotted lines) surfaces are plotted.  }
\label{compareexp1}
\end{figure}

It has been reported that surface roughness will induce an additional threshold energy in quantum well devices and this effect had been systematically measured in experiments \cite{tsuitsui05,uchida03}, i.e. enhanced threshold energy. Therefore, the observed threshold energy will be greater than that described by the more commonly understood body quantization effect according to $\hbar^{2}\pi^{2}/(2m_{z}T_{B}^{2})$. Fig. \ref{compareexp1} compares the theoretically calculated threshold energy of a roughened channel (i.e. perfectly correlated and anti-correlated surfaces morphology) with the available experimental data for Si$\left\langle 110\right\rangle$ and Si$\left\langle 100\right\rangle$ quantum well devices \cite{tsuitsui05,uchida02}. Assuming a roughness amplitude of $A_{0}=0.5nm$ and $A_{0}=0.6nm$ for the Si$\left\langle 110\right\rangle$ and Si$\left\langle 100\right\rangle$ devices respectively, we are able to match the experimental data for the range of quantum well thicknesses $T_{B}$. A roughness wavelength of $\lambda=2.5nm$ was assumed for all devices in our calculations. The corresponding threshold energy due to an unroughened channel are also shown (plotted as solid lines). A quantization mass ($m_{z}$) of $0.9m_{0}$ is employed for the Si$\left\langle 100\right\rangle$ devices. For Si$\left\langle 110\right\rangle$, the effective mass tensor has off-diagonal terms in the direction normal to the quantum film surface. A unitary transformation is employed to decoupled them, as described by Stern and Howard \cite{stern67}. Eventually, a quantization mass of $0.33m_{0}$ is obtained for Si$\left\langle 110\right\rangle$ devices.

The enhanced threshold energy is a geometrically derived property of $2D$ quantization effects. The $2D$ geometry of the roughened channel introduces additional lateral confinement which serves to enhance overall threshold energy. Fig. \ref{resonanceLDOSa} and \ref{resonanceLDOSp} illustrates this lateral confinement effect in channels with perfectly and anti-correlated surfaces. Note that these channels are constructed such that they retain the same volume as the unroughened channel. Eq. \ref{omegacall} describes the approximate enhanced threshold energy due to surface roughness for the case of channels with anti-correlated surface morphology. Eq. \ref{omegacall} tells us that the enhanced threshold energy is related to the material parameters according to $(m_{x}m_{z})^{-1/2}$. Furthermore, contrary to the $T_{B}^{-2}$ dependency in the usual case of body quantization, the enhanced threshold energy exhibits a $T_{B}^{-3/2}$ dependency as shown in Eq. \ref{omegacall}. From the viewpoint of device performances, this translates to an additional threshold voltage shift. A small device threshold voltage shift is an important criterion to suppress the on-chip device-to-device electrical properties variations \cite{low05}. 


\section{\label{sec:level2}Conclusion}
The study of the impact of surface roughness on electronic transport had usually focused on the dissipative regime, with transport dynamics well governed by the classical Boltzmann's transport equation. Electronic transport through a channel with roughened surfaces in the phase coherent regime is less understood. Using the Keldysh non-equilibrium Green function approach within a FEM-BEM numerical scheme, we performed a systematic study of quantum transport through a deca-nanomater length quantum well channel with perfectly correlated and anti-correlated surfaces. Due to the pseudo-periodicity in these simulated structures, their energy-resolved transmission possesses pseudo-bands, pseudo-gaps and an enhanced threshold energy. Channels with perfectly correlated surfaces exhibit wider pseudo-bands than their anti-correlated counterparts. Perfectly correlated channels also permit odd-to-even mode transition, which is not allowable in channel with anti-correlated surfaces. 

An effective transmittivity in these structures is derived, by computing the average transmission over the range of energy within the pseudo-band. By surveying channels with various material parameters combinations (i.e. $m_{x}$ and $m_{z}$), we found that a smaller transport mass $m_{x}$ is beneficial for the transmittivity of the channel and serves to increase the energy bandwidth of the pseudo-band. The observation of the contrasting trends in the dependence of transmittivity on $m_{z}$ for anti-correlated and perfectly correlated surfaces is interesting. A quantum well with perfectly correlated surface is more optimal for channel material with smaller $m_{z}$. Technically speaking, a sufficiently correlated surface can be engineered via techniques like atomic layer deposition. Recall that the `classical' perturbing Hamiltonian due to surface roughness can be attributed to a local energy level fluctuation and a local fluctuation of charge density \cite{mou00}. On a general note, one could then say that channels with anti-correlated surfaces emphasize the former scattering mechanism, while the perfectly correlated surfaces emphasize the latter mechsmism.

Lastly, we studied the phenomenon of enhanced threshold voltage shifts. Excellent corroboration with the experimental data was obtained. Enhanced threshold voltage shifts exhibits a $T_{B}^{-3/2}$ dependency and its contribution to the total threshold energy is significant in the small $T_{B}$ regime. Therefore, suppression of the device-to-device threshold voltage variations in quantum well channels with small quantization mass will present considerable challenge for the semiconductor device industry \cite{itrs07}.

\begin{acknowledgments}
T. Low would like to thank M. S. Lundstrom for suggesting the examination of energy-resolved transmission for a longer channel and T. Manz for proofreading this manuscript. T. Low gratefully acknowledge the support of the Singapore Millenium Post-doctoral Fellowship for the initial support of this work. Also the Network for Computational Nanotechnology and the Nanoelectronics Research Initiative for current support.\\

\end{acknowledgments}


\end{document}